\documentclass[manuscript]{aastex61}

\newcommand\aastex{AAS\TeX}

\usepackage{amsmath}

\received{}
\revised{}
\accepted{}
\submitjournal{ApJL}

\shorttitle{\aastex\ sample article}
\shortauthors{Guo et al.}

\begin{document}

\title{Fast Sausage Oscillations in Coronal Loops with Fine Structures}

\correspondingauthor{Mingzhe Guo}
\email{m.guo@sdu.edu.cn}

\author{Mingzhe Guo}
\affiliation{Shandong Provincial Key Laboratory of Optical Astronomy and Solar-Terrestrial Environment, Institute of Space Sciences, Shandong University, Weihai 264209, China}

\author{Bo Li}
\affiliation{Shandong Provincial Key Laboratory of Optical Astronomy and Solar-Terrestrial Environment, Institute of Space Sciences, Shandong University, Weihai 264209, China}

\author{Mijie Shi}
\affiliation{Shandong Provincial Key Laboratory of Optical Astronomy and Solar-Terrestrial Environment, Institute of Space Sciences, Shandong University, Weihai 264209, China}
\affiliation{Centre for mathematical Plasma Astrophysics, Department of Mathematics, KU Leuven, 3001 Leuven, Belgium}




\begin{abstract}

Fast sausage modes (FSMs) in flare loops have long been invoked to account for rapid quasi-periodic pulsations (QPPs) with periods of order seconds in flare lightcurves. 
However,
most theories of FSMs in solar coronal cylinders assume a perfectly 
axisymmetric equilibrium,
an idealized configuration apparently far from reality.
In particular,
it remains to examine whether FSMs exist in coronal cylinders with fine structures.
Working in the framework of ideal magnetohydrodynamics (MHD),
we numerically follow the response to an axisymmetric perturbation of a coronal cylinder for which a considerable number of randomly distributed fine structures are superposed on an axisymmetric background.
The parameters for the background component are largely motivated by the recent IRIS identification of a candidate FSM in Fe XXI 1354 \AA~observations.
We find that the composite cylinder rapidly settles to an oscillatory behavior largely compatible with a canonical trapped FSM.
This happens despite that kink-like motions develop in the fine structures.
We further synthesize the Fe XXI 1354 \AA~emissions,
finding that the transverse Alfv\'en time characterizes the periodicities in the intensity,
Doppler shift,
and Doppler width signals.
Distinct from the case without fine structuring,
a non-vanishing Doppler shift is seen even at the apex.
We conclude that density-enhanced equilibria need not be strictly axisymmetric to host FSM-like motions in general,
and FSMs remain a candidate interpretation for rapid QPPs in solar flares.

\end{abstract}


\section{INTRODUCTION} 
\label{sec_intro}

Rapid quasi-periodic pulsations (QPPs) with periods ranging from seconds to a couple of tens of seconds are frequently seen in solar flare light curves measured in various passbands \citep[see reviews by][]{2009SSRv..149..119N, 2016SoPh..291.3143V, 2020SSRv..216..136L,2021SSRv..217...66Z}.
A possible interpretation for these rapid QPPs is associated with fast sausage modes (FSMs) supported in flare loops.
The FSMs manifest axisymmetric oscillating property 
under a classical assumption that the magnetic waveguides are considered as straight axisymmetric cylinders \citep[][]{1983SoPh...88..179E, 1984ApJ...279..857R}. 
Divided by cutoff wavenumbers,
the FSMs possess two regimes:
wave energy is well confined in the trapped regime but is continuously lost to the surroundings when the leaky regime arises.
In both regimes,
the periods of the FSMs are found to be determined by the transverse Alfv\'en transit time,
which typically evaluates to seconds in the corona.
This makes the FSMs a potential candidate to account for the rapid QPPs in solar flares.

A departure from the perfectly axisymmetric straight equilibrium is more reasonable in the structured solar atmosphere.
For simplicity,
we insist on straight equilibria throughout.
A straight cylinder with an elliptical cross-section is a natural consideration that breaks the axisymmetry \citep[e.g.,][]{2003A&A...409..287R, 2011A&A...527A..53M,2020ApJ...904..116G}.
Actually,
many pores and sunspots are measured to have elliptical cross-sections \citep[e.g.,][]{2018ApJ...857...28K, 2021ApJ...912...50A},
and FSMs have been proved to be supported in elliptical cylinders both theoretically \citep[][]{2009A&A...494..295E, 2021ApJ...912...50A} and observationally \citep[e.g.,][]{2018ApJ...857...28K}.
Equilibria with more realistic irregular cross-sections have been discussed in \citet[][]{2021thesis_A}.
Furthermore,
recent observations by the Interface Region Imaging Spectrograph (IRIS) have revealed that FSMs are supported by fine-structured flare loops \citep[][hereafter {T16}]{2016ApJ...823L..16T}.
As typical magnetic structures in the solar corona,
coronal loops are generally not monolithic in realistic measurements,
but consist of fine sub-structures instead \citep[e.g.,][for a review]{2017ApJ...840....4A,2014LRSP...11....4R}.
The investigations associated with waves and oscillations in loops with fine structures have attracted substantial attention,
such as eigenmode analysis in two parallel loops \citep{2008ApJ...676..717L, 2008A&A...485..849V, 2010A&A...515A..33R, 2014A&A...562A..38G}
and studies of transverse oscillations in more complex multistranded loop systems with application of the T-matrix theory \citep{2010ApJ...716.1371L, 2019A&A...629A..20L},
 which was first introduced to the solar context to study p-mode in sunspots by \citet[][]{1991ApJ...379..758B} and \citet[][]{1994ApJ...436..372K}.
In addition,
transverse waves or oscillations have been examined from the initial value problem perspective in multistranded loops \citep{2008ApJ...679.1611T, 2016ApJ...823...82M, 2019ApJ...883...20G}.
However,
these studies focus on transverse waves or oscillations.
Regarding FSMs,
a forward step has been made by considering concentric shells as the radial inhomogeneities of straight magnetic waveguides \citep{2007SoPh..246..165P, 2015SoPh..290.2231C}.
However,
the axisymmetry remains.

To our knowledge,
there is no study on FSMs in a non-monolithic loop with fine structures so far and the existence of sausage modes in such kind of equilibrium remains unknown.
We thus perform a three-dimensional MHD simulation involving FSMs in a composite loop with randomly distributed fine structures inside.
This paper is organized as follows.
Section \ref{sec_models} details the
loop model we considered,
including the equilibrium and numerical setup.
In Section \ref{sec_results} we present the results and forward modelling.
Section \ref{sec_conclusion} summarizes the results,
ending with some discussion.

\section{NUMERICAL MODEL}
\label{sec_models}
\subsection{Equilibrium Setup}
\label{sec_sub_IC}

\begin{figure}
	\centering
	\gridline{\fig{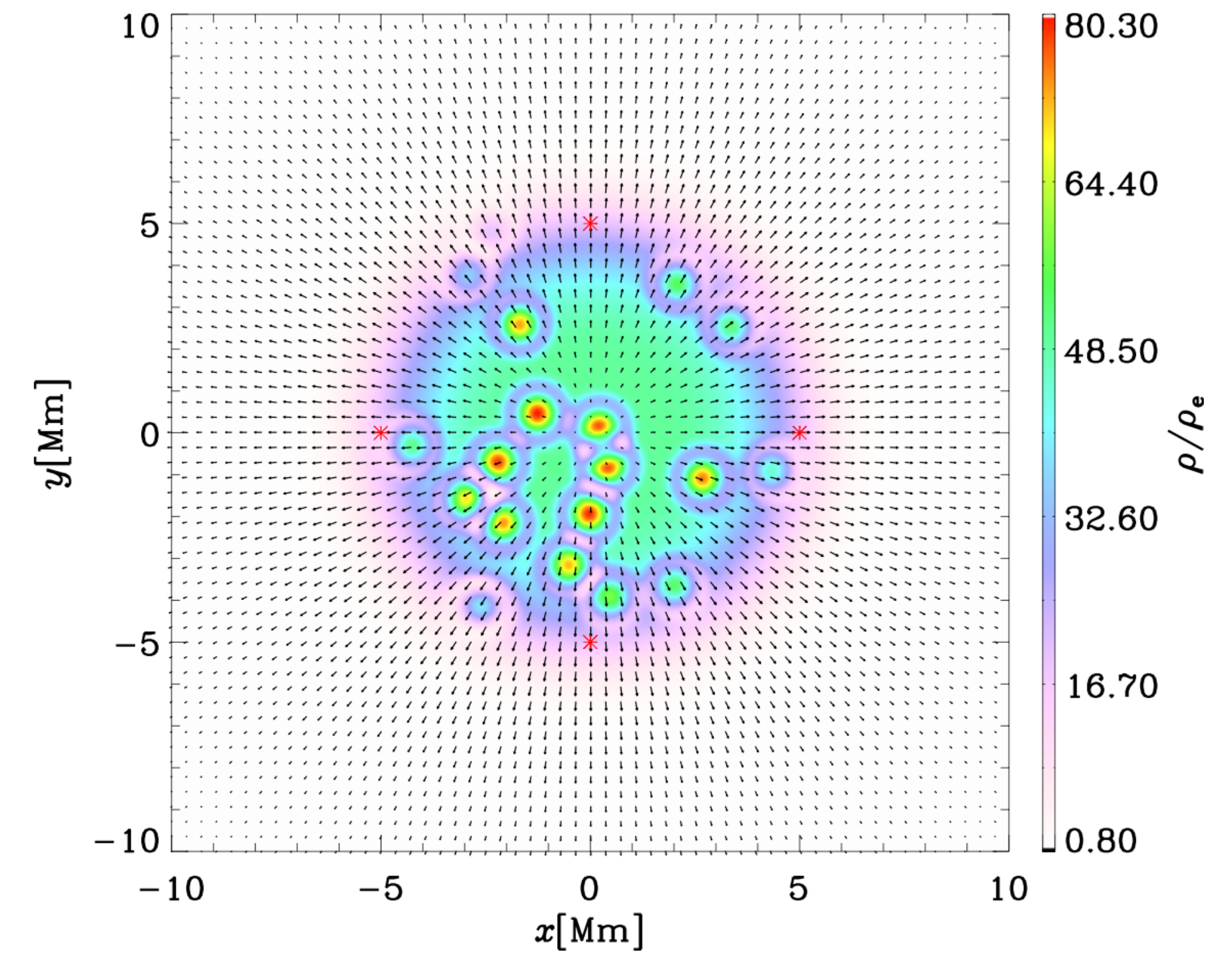}{0.8\textwidth}{}
	}
	\caption{
	   Initial density snapshot of loop cross-section at loop apex ($z=L/2$).
       The density is in the unit of background density $\rho_{\rm e}$.
       Black arrows represent the initial velocity field. 
       Red asterisks denote four representative positions analyzed in the following. The snapshot is taken from the animation attached to this figure. The animation evolves from $t=0$ to $t=300$s. The duration is $30$s.}
			
	\label{fig_snapshot}
\end{figure}

We consider a loop model as a monolithic cylinder with fine structures randomly spread inside.
Note that this setup is our first attempt towards the non-axisymmetric equilibrium,
it would be better to retain the monolithic background for reference.
The parameters of the loop are uniform along the vertical direction.
In the transverse direction,
the density profile is given by
\begin{eqnarray}
\rho(x, y) = \rho_{\rm mono}(x,y)+\rho_{\rm FS}(x,y)~,
\label{eq_rhoEq}
\end{eqnarray} 
where $\rho_{\rm mono}(x,y)$ and $\rho_{\rm FS}(x,y)$ represent density profiles of the monolithic background and the fine structures, respectively.
They are prescribed by
\begin{eqnarray}
&& \rho_{\rm mono} = \rho_{\rm e}+(\rho_{\mathrm i}-\rho_{\mathrm e})f(x, y)~,
     \label{eq_rhoMono}\\
&& \rho_{\rm FS} = (\rho_{\mathrm i}-\rho_{\mathrm e})f(x, y)g(x,y)~,
     \label{eq_rhoFS}
\end{eqnarray} 
where
\begin{eqnarray}
f(x,y) = \exp\left[-\left(\frac{r}{R}\right)^{\alpha}\right]~,
     \label{eq_rho_f}
 \end{eqnarray} 
with 
\begin{eqnarray}
&& r = \sqrt{x^2+y^2}~,
    \label{eq_rho_r}
 \end{eqnarray} 
 and
\begin{eqnarray}
g(x,y) = \frac{\displaystyle\sum_{j=1}^{N_{\rm FS}} \left[\exp\left(-\bar{r}^{\alpha}_j\right)\cos\left(\pi\bar{r}_j\right)\right]}{\left|\displaystyle\sum_{j=1}^{N_{\rm FS}} \left[\exp\left(-\bar{r}^{\alpha}_j\right)\cos\left(\pi\bar{r}_j\right)\right]\right|_{\rm max}}~~,
     \label{eq_rho_g}
\end{eqnarray} 
with 
\begin{eqnarray}
\bar{r}_j = \frac{\sqrt{(x-x_j)^2+(y-y_j)^2}}{R_{\rm FS}}~.
    \label{eq_rho_rbar}
\end{eqnarray} 
We consider an electron-proton plasma throughout. The mass density $\rho$ is then connected to the electron number density $N$ through $\rho = N m_p$ with $m_p$ being the proton mass. We specify the internal loop density $\rho_{\rm i}$ and external loop density $\rho_{\rm e}$ in Equation \eqref{eq_rhoMono} such that the corresponding $N$ is $5.0\times10^{10}{\rm cm}^{-3}$ and $0.8\times10^9 {\rm cm}^{-3}$, respectively. 
The loop length is fixed at $L=45 {\rm Mm}$,
and the nominal loop radius is $R=5 {\rm Mm}$.
The radius of each fine structure is $R_{\rm FS}=0.8 {\rm Mm}$.
The steepnesses of density profiles of the monolithic background and the fine structures are determined by a parameter of $\alpha=5$.
$\left[x_j, y_j\right]$ in Equation \eqref{eq_rho_r} represents the position of the center of each fine structure,
which is supposed to be random in the monolithic region ($\left|x,y\right|\le R$).
$N_{\rm FS}$ represents the number of fine structures,
we take $N_{\rm FS}=20$ in practical runs.
Figure \ref{fig_snapshot} shows the initial snapshot of density distribution at $z=L/2$.
In addition,
the distribution of temperature follows the same profile of the density of the monolithic background.
We take the temperature inside the loop as $T_{\rm i}=10{\rm MK}$ and external temperature as $T_{\rm e}=2{\rm MK}$.
Furthermore,
to maintain magnetostatic pressure balance, 
the magnetic field has a variation from $B_{\rm i}=50{\rm G}$ in the internal region of the monolithic cylinder to $B_{\rm e}=77.3{\rm G}$ in the external medium.
The resulting internal (external) Alfv{\'e}n speed is $v_{\rm Ai}=496 {\rm km~s^{-1}}$ ($v_{\rm Ae}=5965 {\rm km~s^{-1}}$).
The physical and geometrical parameters of the monolithic component follow rather closely from the IRIS measurements of the composite flare loop supporting a candidate FSM in T16.

\subsection{Numerical Setup}
\label{sec_sub_setup}

Trying to excite FSMs in the composite loop,
we adopt an axisymmetric initial velocity perturbation,
which is similar to the one used in \citet{2016ApJ...833..114C},
\begin{eqnarray}
\delta v_x(x,y,z; t=0)= v_0\frac{r}{\sigma_r}\exp\left[\frac{1}{2}\left(1-\frac{r^2}{\sigma^2_r}\right)\right]\sin\left(\frac{\pi z}{L}\right)\left(\frac{x}{r}\right)~,
     \label{eq_v0x}
 \end{eqnarray} 
 \begin{eqnarray}
\delta v_y(x,y,z; t=0)= v_0\frac{r}{\sigma_r}\exp\left[\frac{1}{2}\left(1-\frac{r^2}{\sigma^2_r}\right)\right]\sin\left(\frac{\pi z}{L}\right)\left(\frac{y}{r}\right)~,
     \label{eq_v0y}
 \end{eqnarray} 
where $v_0=10{\rm km}~{\rm s}^{-1}$ is the amplitude of the initial velocity.
$\sigma_r = 5.0 {\rm Mm}$ characterizes the extent of the perturbation in the radial direction.
 See Figure \ref{fig_snapshot} for the initial velocity field (black arrows) at $z=L/2$.
Note that our initial velocity perturbation is not intended to represent a realistic exciter.
Using this initial perturbation is computationally simple and can readily compare with the results in monolithic loops.

To solve the three-dimensional ideal MHD equations,
we use the PLUTO code \citep{2007ApJS..170..228M}.
A piecewise parabolic scheme is employed for spatial reconstruction.
The numerical fluxes are computed by the HLLD Riemann solver,
and the second-order Runge-Kutta algorithm is used for time marching.
A hyperbolic divergence cleaning method is adopted to maintain the divergence-free condition of the magnetic field.
The computation domain is $[-10,10]{\rm Mm} \times [-10,10]{\rm Mm} \times [0,L]$.
We employ a uniform grid of 64 points from 0 to $L$ in the $z$-direction,
and 840 uniformly spaced cells in the $x$- and $y$-direction, respectively.

The boundary conditions are specified as follows.
We fix the transverse velocities at both ends of the loop to be zero,
while $v_z, B_x, B_y$ are set to have zero-gradients.
The other variables are fixed as their initial values.
Outflow conditions are used in all the lateral boundaries.

\section{RESULTS}
\label{sec_results}

\subsection{Oscillations in the Composite Loop}
\label{sec_sub_resultsWaves}

Inspired by the IRIS observation in T16,
we excite axial fundamental sausage modes in the composite loop,
meaning that the axial wavenumber is $\pi/L$. 
Recall that the ratio between the loop length and radius is $L/R=9$, and the steepness of the transverse structuring of the monolithic loop is $\alpha=5$,
we can readily find that the normalized cutoff wavenumber in the monolithic background without fine structures is $k_cR=0.32$,
by solving pertinent eigenvalue problem in monolithic loops.
The cutoff wavenumber is smaller than the axial wavenumber,
meaning that sausage modes in the monolithic loop would be in the trapped regime.
For a reference,
we can also estimate the period of the FSM being $16.4$s in the monolithic loop.

\begin{figure}
	\centering
	\gridline{\fig{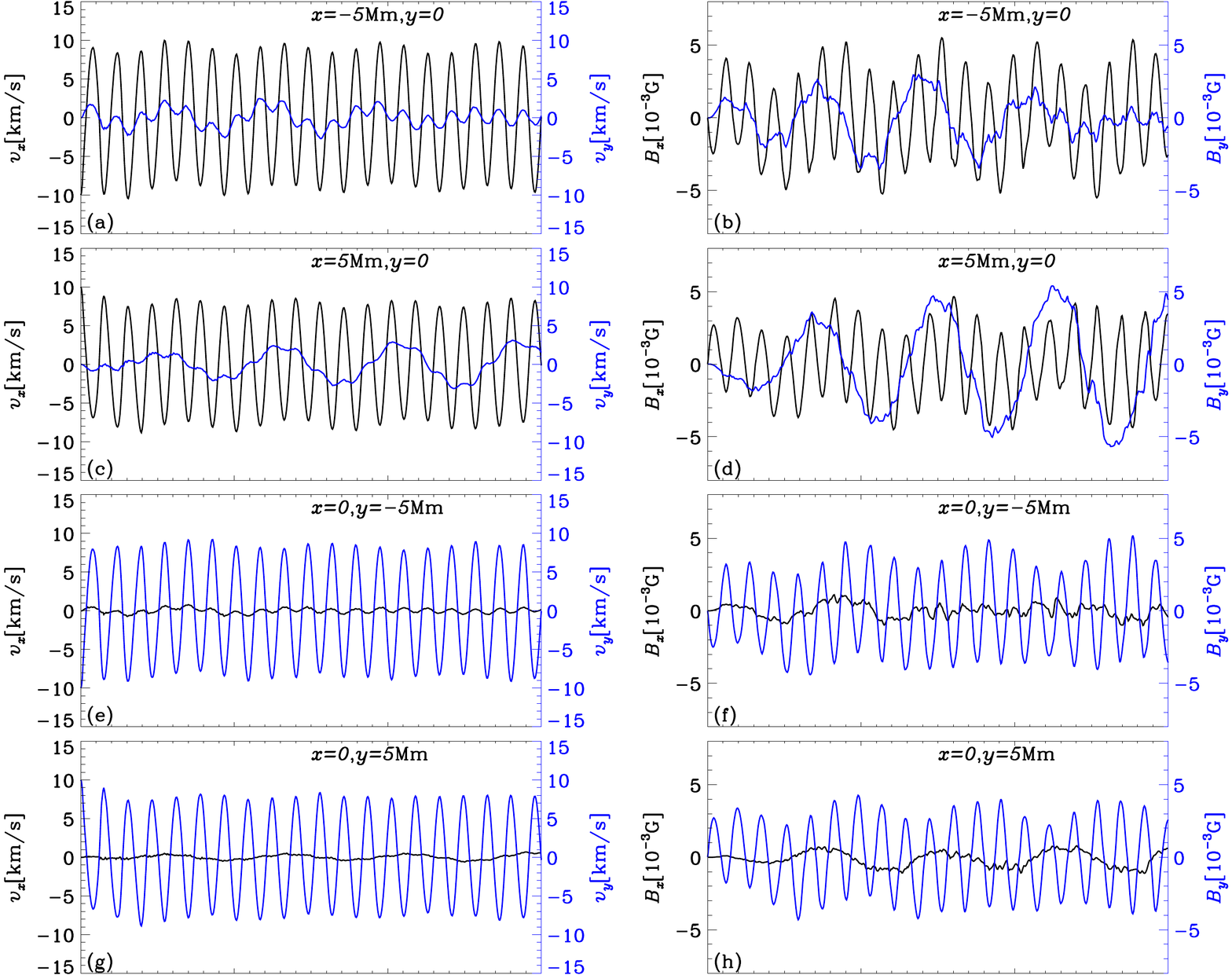}{0.9\textwidth}{}
	}
	\caption{
	  Left: temporal evolution of $v_x$ (black lines) and $v_y$ (blue lines)  sampled at four representative positions as labelled at $z=L/2$.  Right: similar to the left column but for the magnetic field. }			
	\label{fig_velocity}
\end{figure}

Now we examine the oscillations in the composite loop after the initial velocity perturbation.
Figure \ref{fig_velocity} presents the profiles of velocity and magnetic field at four representative locations at the loop boundary,
namely $[x,y,z] = \{[-R,0,L/2],[R,0,L/2],[0,-R,L/2],[0,R,L/2]\}$.
The oscillation profiles of $v_x$ at $y=0$ and $v_y$ at $x=0$ show no significant damping,
indicating a trapped mode.
Comparing Figure \ref{fig_velocity} (a) and (c) (or Figure \ref{fig_velocity} (e) and (g)),
we find that $v_x$ ($v_y$) at $y=0$ ($x=0$) are out of phase.
Meanwhile,
the amplitudes of $v_x$ at $x=0$ are relatively small.
Similar properties can also be found in the magnetic field curves.
All the above properties indicate a typical breathing-like motion in the whole loop
since the four sample points are chosen axisymmetric.
We find the period of the oscillation is about 15s,
which is close to the predicted period of an FSM in the monolithic loop without fine structures.
We thus say that the current composite loop supports an FSM,
despite that the loop is not axisymmetric with the appearance of randomly distributed fine structures therein.
Note that the $v_y$ curves in Figure \ref{fig_velocity} (a) and (c) present relatively long-time-scale periodicity.
The period is very close to the local Alfv\'en period ($\sim77.9$s),
 indicating the excitation of Alfv\'en waves around fine structures.
A phase difference of $\pi/2$ can be observed between the $B_y$ and $v_y$ profiles at $[x,y,z] = [\pm R,0,L/2]$.
This is a typical signature of standing Alfv\'en waves (see \citeauthor{2020ApJ...904..116G}~\citeyear{2020ApJ...904..116G}).
The Alfv\'en waves can be clearly observed near the fine structures at the loop boundary at $y=0$,
hinting that these Alfv\'en waves originate from the oscillations of these fine structures.

\begin{figure}
	\centering
	\gridline{\fig{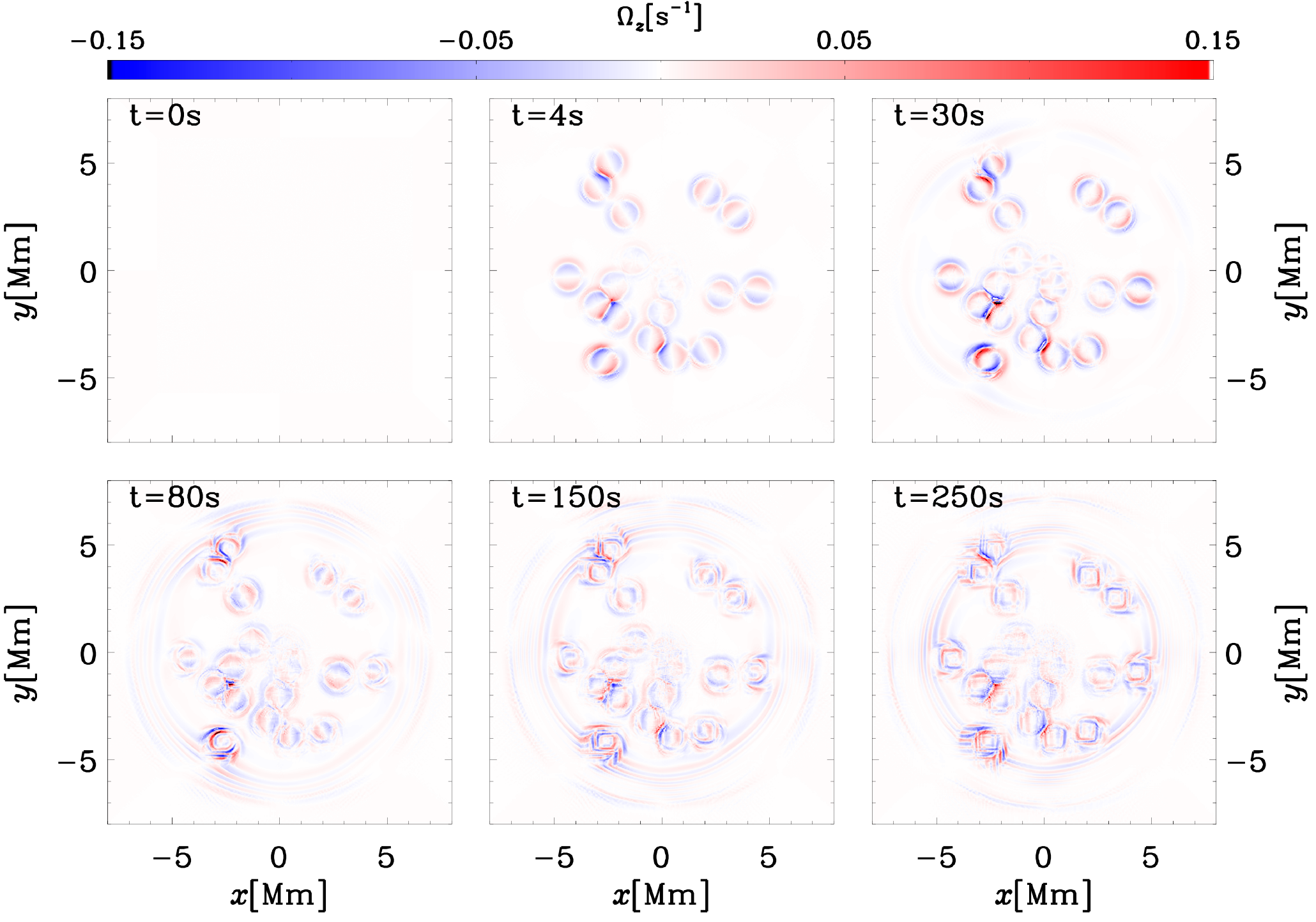}{0.8\textwidth}{}
	}
	\caption{
	   	$z$-component of the vorticity at $z=L/2$ at different instants.  The snapshots are taken from the animation of the $\Omega_z$ attached to this figure. The animation evolves from $t=0$ to $t=300$s. The duration is $30$s.
	}			
	\label{fig_vort}
\end{figure}

\begin{figure}
	\centering
	\gridline{\fig{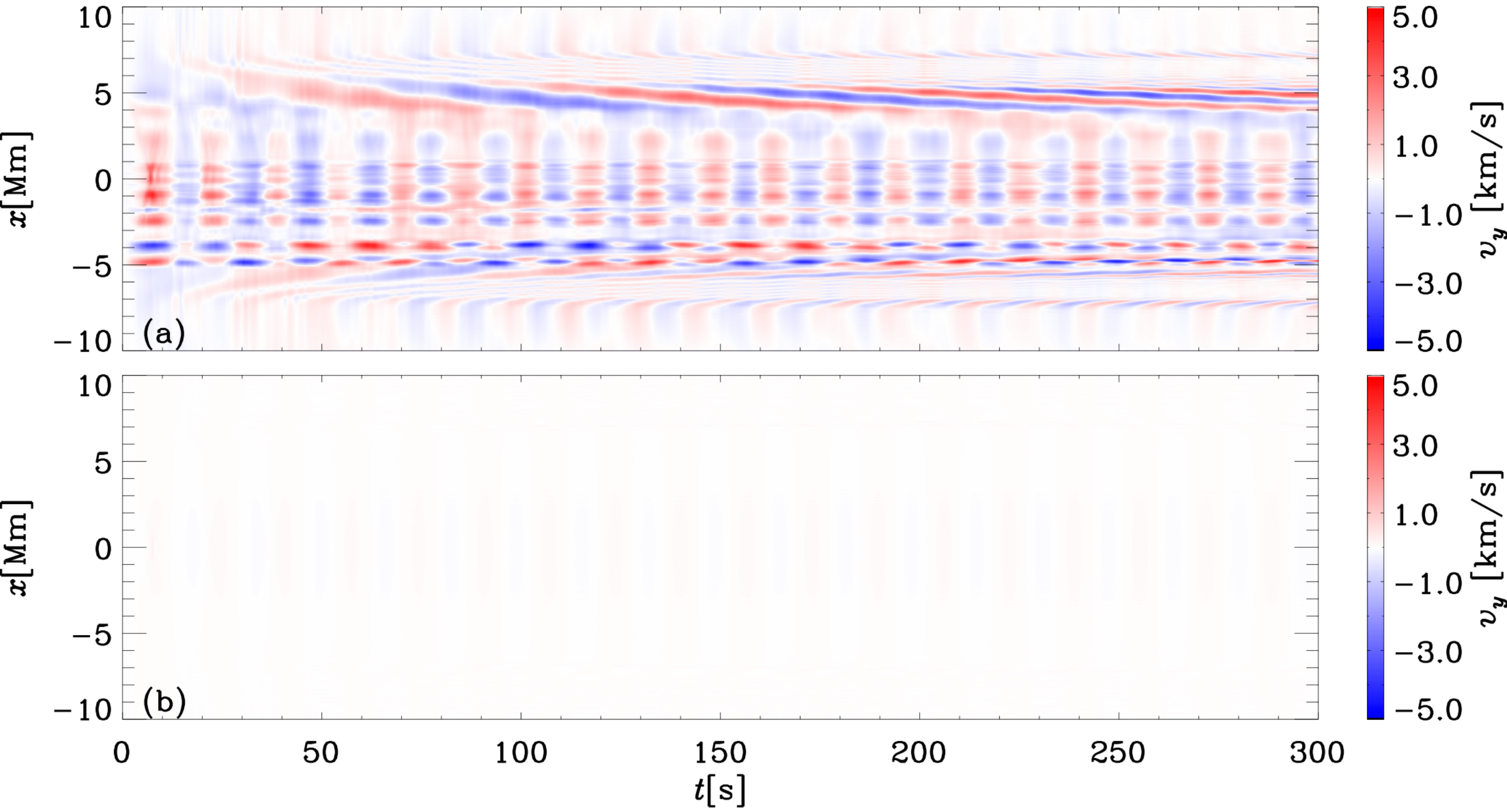}{0.9\textwidth}{}
	}
	\caption{
	   	(a) Temporal evolution of $v_y$ at $z=L/2$. (b) Similar to (a) but in a monolithic loop with no fine strands.
	}			
	\label{fig_vy}
\end{figure}

The dynamics of fine structures can be clearly revealed by examining the $z$-component of vorticity ($\Omega_z$) at different instants,
as shown in Figure \ref{fig_vort}.
The vorticity is zero at the initial state,
showing no vortex of the initial perturbation.
We find that the internal regions of fine structures possess an opposite oscillating direction against their boundary,
indicating kink-like motions that are rapidly excited by the initial perturbation.
Similar properties can also be found in the velocity field of kink modes \citep[e.g.,][]{2014ApJ...788....9G,2020ApJ...904..116G}.
Resonantly converted Alfv\'en waves are characterized by blue and red ripples around each fine structure (see also the associated animation of Figure \ref{fig_vort}). 
Before proceeding,
we examine the temporal evolution of $v_y$ in Figure \ref{fig_vy}(a).
Blue and red ripples near the loop boundary become more and more inclined,
leading to an increase of small structures at a given time. 
A similar property has often been observed in previous studies (see Figure 12 in \citeauthor{2019A&A...631A.105H}~\citeyear{2019A&A...631A.105H} and Figure 4 in \citeauthor{2020ApJ...904..116G}~\citeyear{2020ApJ...904..116G}).
The increasing ripples are usually interpreted as the phase mixing of the localized Alfv\'en waves.
So a process involving resonant conversion from kink modes to local Alfv\'en waves and subsequent phase mixing of Alfv\'en waves is present in the fine structures.
In addition,
we see the distortion of the fine structures as time increases due to the velocity shear between the fine structures and their ambient surroundings.
This distortion is reinforced by the phase mixing,
inducing more substructures.
We also find the appearance of blue and red ripples around the loop boundary after about $t=30$s.
The ripples extend from the boundary to the internal region of the loop.

We see small ripples in the external loop region in the velocity evolution in Figure \ref{fig_vy} (a),
which can also be seen in Figure \ref{fig_vort}.
Although it has been demonstrated that the FSM is in the trapped regime,
small amplitude oscillations can still be observed outside the loop region due to the scattering of the substructures.
To prove this,
we perform a new simulation in the monolithic loop without substructures and plot the $v_y$ profile in Figure \ref{fig_vy} (b).
No oscillating signals nor small structures can be seen in 
Figure \ref{fig_vy} (b),
meaning that the small ripples in the external loop region are indeed from the substructures.

Although the oscillations seem complex in the composite loop,
the predominant wave mode in the loop region is the FSM.
A short-time-scale periodicity of the order of transverse Alfv\'en transit time can be seen in non-axisymmetric  loops with fine structures,
making it possible to interpret the oscillations with periods of several seconds in observations with FSMs.
The FSM remains when considering different numbers and distributions of the strands. 
Only the amplitudes of kink-like motions of the strands get varied as their locations with respect to the loop center vary. 
The resonances can be seen in all cases unless the strands are located at loop center.
In addition,
a lower resolution run with a set of $420\times420\times32$ grid cells has been conducted.
We find no significant difference in the oscillation curves from those in Figure \ref{fig_velocity}. 
Even though fewer ripples around the strands resulting from phase mixing are seen,
the above analysis is not influenced in the lower resolution run.

In fact, 
sausage modes characterized by coherent breathing motions have also been studied in the lower atmosphere,
despite the irregular shape of the examined structures and the fine structuring therein.
 In the context of p-mode interaction with sunspots,
 sausage modes are found in a bundle of symmetric tubes,
 characterized by breathing motions of the tubes \citep[][]{1994ApJ...436..372K}.
 Unlike the FSMs in coronal loops,
 the sausage motions appear in tightly packed sunspot fibrils and are demonstrated as surface modes.

\subsection{Observable Properties}
\label{sec_sub_resultsObserv}

Inspired by the fundamental FSMs measured in flare loops in T16,
we forward-modelled the numerical results by using the FoMo code \citep{2016FrASS...3....4V} to obtain their observational signals.
We perform the synthetic process by using the Fe  \uppercase\expandafter{\romannumeral 21} 1354 \AA~spectral line,
which has been considered by \citet{2019ApJ...874...87S} in their forward modeling effort associated with the above-mentioned IRIS measurements.
In our case,
however,
the coronal loop with substructures therein is a more realistic consideration. The possible influence of  substructures on the detection of FSMs in flare loops is worth examining. 
 We note further that the coronal approximation, inherent to FoMo, 
applies to Fe  \uppercase\expandafter{\romannumeral 21} 1354\AA\ line despite the large values of the electron density
\citep[see][for details]{2019ApJ...874...87S}.

Figure \ref{fig_fomo} illustrates the synthetic intensity,
Doppler velocity and line width at loop apex with a line-of-sight (LoS) along the $y$-axis. 
The intensity shows a periodic enhancement,
indicating the compression induced by the FSM.
This is similar to the monolithic loop results obtained by considering 
the Fe  \uppercase \expandafter{\romannumeral 21} 1354 ${\rm \AA}$ in \citet{2019ApJ...874...87S} and lower temperature lines in \citet{2013A&A...555A..74A}.
Meanwhile,
fine structures are seen all the time in the current intensity variations along the $x$-axis.
Though the blue- and redshifts are disturbed by the fine structures,
the overall periodicity of half the FSM period can be observed in the Doppler shifts.
Recall that the LOS is perpendicular to the $z$-axis in the present model,
we see a net Doppler velocity of $5~{\rm km~s^{-1}}$ due to the broken axisymmetry in the current model.
This has not been reported in previous literature concerning axisymmetric oscillations in a monolithic loop,
in which red- and blueshifts cancel out at loop apex.
This variation in Doppler shift relax the limitation that the LOS should not be perpendicular to a loop axis to obtain a non-zero Doppler velocity.
As for the line width,
a periodicity of half the FSM period shows up again,
illustrating a peanut-like shape,
which has been discussed in \citet{2019ApJ...874...87S}.

\section{DISCUSSION AND CONCLUSIONS}
\label{sec_conclusion}
 
FSMs in flare loops have long been invoked to interpret rapid QPPs with periods of several seconds in flare lightcurves. 
Recent IRIS observations manifest FSMs in fine-structured flare loops.
We thus model a monolithic flare loop with randomly distributed fine structures therein.
By considering an axisymmetric perturbation,
we find that the composite cylinder rapidly settles to a trapped FSM.
A forward model based on the numerical results is further obtained,
showing the observable properties of the FSM in the Fe XXI 1354 \AA~spectral line.
The periodicities characterized by the Alfv\'en transit time can be observed in the intensity,
Doppler shift,
and line width signals.
A periodic Doppler shift can be seen at the loop apex,
illustrating a significant difference from the equilibrium without fine structuring.
In general,
magnetic equilibria need not be perfectly axisymmetric to support FSMs,
which remain a candidate interpretation for rapid QPPs in solar flares.

The short periodic signals that appear in the non-axisymmetric substructured loop relax the strict definition of FSMs in coronal loops.
In the corona,
complex density distribution such as substructures in flare loops is still an observational obstacle for unambiguous identifications and measurements of FSMs.
Our current results thus bring more confidence to the identification of FSMs in such kind of complex loop structure and the interpretation of rapid QPPs.

In reality,
flare loops are more complex and dynamic than the current model.
Chromospheric evaporation has been reported in the impulsive phase in T16.
Simulations by \citet[][]{2019ApJ...877L..11R} demonstrated that 
turbulent interactions may be induced by the chromospheric evaporation in a magnetic arcade,
leading to
the fast mode associated QPPs in soft X-ray.
So both observations and simulations indicate a close relation between the evaporation and QPPs.
However,
this process cannot be evaluated in our current model.
Although the curvature and many realistic processes have not been included,
we stress again that the current cylinder aims to achieve a conceptional understanding of the existence of FSMs in the non-axisymmetric structure,
rather than a full picture of a flare event.
Realistic information has been considered in recent models \citep[e.g.,][]{2019NatAs...3..160C,2020ApJ...896...97R}, where X-ray lightcurves that are used for quantifying QPPs are obtained.

Another realistic information in flare loops is gravitational stratification.
Including the stratification would induce a density reduction from loop top to bottom.
The density scale height,
which can be roughly estimated by $k_B T_{\rm i}/\mu m_{\rm p} g$ with $k_B$ the Boltzmann constant and $\mu$ the mean molecular weight,
would be 500 Mm. 
This value is much larger than the loop height ($L/\pi\sim14$Mm) when deformed to a semi-circle. 
So the density would change only $3\%$ from the loop top to bottom in the current model. 
Once the stratification is considered,
the Rayleigh-Taylor dynamics that are believed to be important in prominences \citep[e.g.,][]{2015ApJ...806L..13K, 2015ApJ...799...94T} is worth noticing at loop apex.
In the current model, 
the largest growth rate of the Rayleigh-Taylor instability can be roughly obtained by $\gamma=\sqrt{g \pi /R}\sim0.01{\rm s^{-1}}$ \citep[][and references therein]{2019book_Goedbloed}.
This means that
the growth time of the Rayleigh-Taylor instability is much larger than the oscillation period.

The origins of QPPs are attributed to various mechanisms.
Apart from the aforementioned 
chromospheric evaporation,
repetitive magnetic reconnection can account for long-period QPPs \citep[e.g.,][]{2006ApJ...644L.149O}. 
MHD oscillation is also a possible candidate to account for these QPPs, as demonstrated in T16.
Using the FSM to interpret the the phase-difference between the intensity and Doppler shift seems more reasonable,
and this interpretation is supported by forward models \citep[][]{2013A&A...555A..74A}.

One caveat of the present model might be the composite loop seems artificial.
As a first step towards the non-axisymmetry,
a monolithic loop with substructures is worth considering
since the well-known results in monolithic loops \citep[e.g.,][]{2016SoPh..291..877G, 2019ApJ...874...87S} can be a reference for the new findings.
Besides,
it would not be easy to unambiguously exclude the possibility of substructures interspersing a monolithic loop.
Fine structures can also be induced by
transverse oscillations in a monolithic loop \citep[e.g.,][]{2014ApJ...787L..22A}.
Although the pre-existing fine structures are different from the transverse wave-induced ones,
the similar strand-like structures manifested in forward models indicate that the two scenarios are not easy to distinguish. 
In addition,
fine structures play a role to break the axisymmetry of the monolithic background.
Although this can be readily achieved by considering a single strand inside the monolithic loop,
a considerable number of strands are needed
\citep[e.g.,][]{2013A&A...556A.104P}.

\acknowledgments
 {The authors acknowledge the funding from the National Natural Science Foundation of China (41974200, 11761141002, 41904150). }

\clearpage
 \appendix

\section{Oscillation Profiles in a Cylindrical Coordinate}
\label{sec_appendix_A}

 The radially and azimuthally averaged evolutions in velocity and magnetic field along the loop boundary at $z=L/2$ are shown in Figure \ref{fig_vel_app}.
The quantity $q$ are calculated by 
\begin{eqnarray}
\left\langle q\right\rangle=\frac{1}{2\pi R}\int_l q dl~,
    \label{eq_ave}
\end{eqnarray} 
where $q$ represents the radial (azimuthal) velocity $v_r$ ($v_\phi$) and
the radial (azimuthal) magnetic field $B_r$ ($B_\phi$).
$l$ is the length along the loop edge at $z=L/2$.
A similar oscillation property as in the main text can be seen in both $\left\langle v_r \right\rangle$ and $\left\langle B_r \right\rangle$ profiles. 
A trapped FSM with a period of about 15s shows up.
However,
the Alfv\'en signal in $\left\langle v_\phi \right\rangle$ and $\left\langle B_\phi \right\rangle$ evolutions can hardly be seen since the Alfv\'en waves appear locally near the fine strands and are naturally hidden by the averaged calculation.

\section{Alfv\'en Resonance Examination}
\label{sec_appendix_B}

The density variation in the substructures inside the loop leads to local Alfv\'en resonance.
Figure \ref{fig_res} shows the strength of local velocity $v=\sqrt{v_x^2+v_y^2}$ at loop apex at different time. The largest values of the velocity in each panel are larger than the amplitude of initial perturbation ($10{\rm km s^{-1}}$),
and their locations correspond to the positions of local resonance \citep[see also Fig.5 in][]{2020A&A...636A..40H}.
Local Alfv\'en frequency $\omega_A=\pi v_{\rm A}/L$ is also contoured.
We see the resonance happens at about $0.05{\rm rad~s^{-1}}\leq \omega_A\leq0.13 {\rm rad~s^{-1}}$,
this range roughly outlines the locations of the largest values of the velocity around each strand.
The Alfv\'en frequency in this range is between the internal and external one.
Comparing with Figure \ref{fig_vort} we find that the resonance locations are right around the places where the fine structures present kink-like motions.
So the resonant conversion from kink motions to local Alfv\'en waves in the fine strands is proved.
Note that the Alfv\'en signals in Figure \ref{fig_velocity} shows a period of about $77.9$s,
which corresponds to the local Alfv\'en frequency of $0.08{\rm rad~s^{-1}}$ at $[x,y]=[5{\rm Mm},0]$ in Figure \ref{fig_res}.

\section{Observable Properties in Another LOS Orientation}
\label{sec_appendix_C}

\begin{figure}
	\centering
	\gridline{\fig{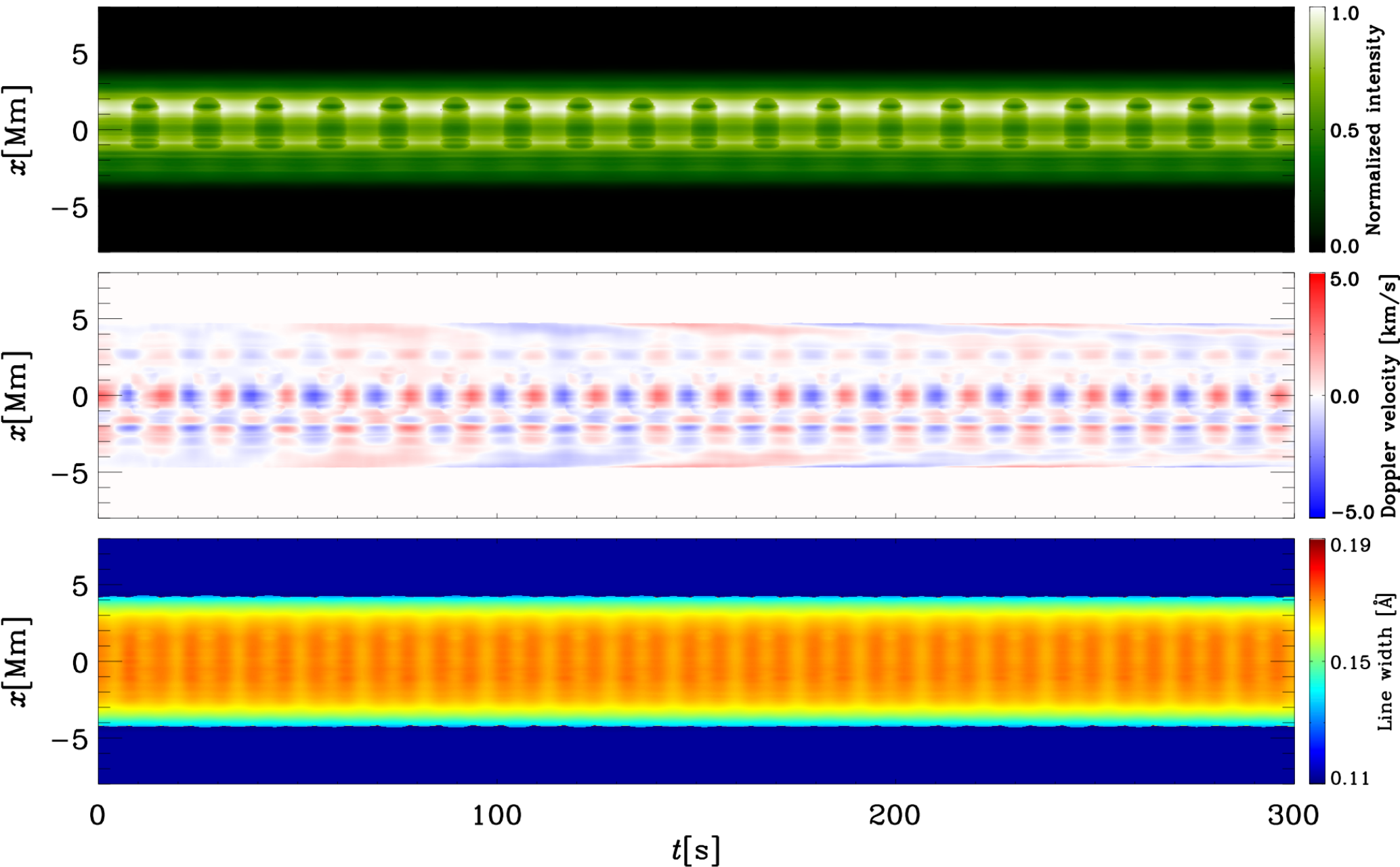}{0.8\textwidth}{}
	}
	\caption{
	   	Temporal evolution of synthetic intensity, Doppler velocity, and line width in the Fe  \uppercase\expandafter{\romannumeral 21} 1354 \AA~spectral line at $z=L/2$ with an LoS along the $y$-axis. 
	}			
	\label{fig_fomo}
\end{figure}

\begin{figure}
	\centering
	\gridline{\fig{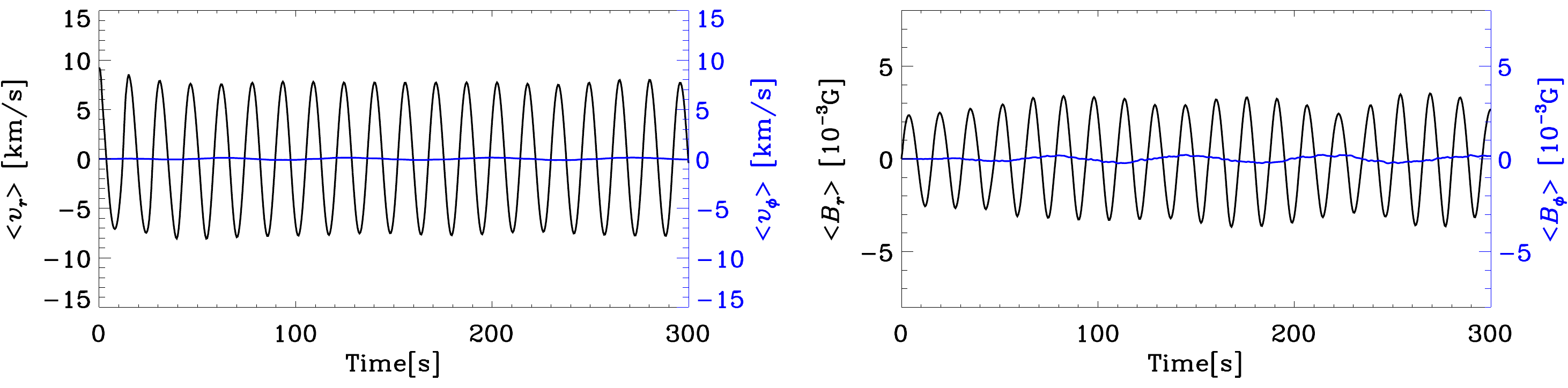}{0.9\textwidth}{}
	}
	\caption{
	   	Left panel shows the radially (azimuthally) averaged velocity evolution $\left\langle v_r \right\rangle$ ($\left\langle v_\phi \right\rangle$) along the loop edge at $z=L/2$. Right: similar to the left column but for the magnetic field.
	}			
	\label{fig_vel_app}
\end{figure}

\begin{figure}
	\centering
	\gridline{\fig{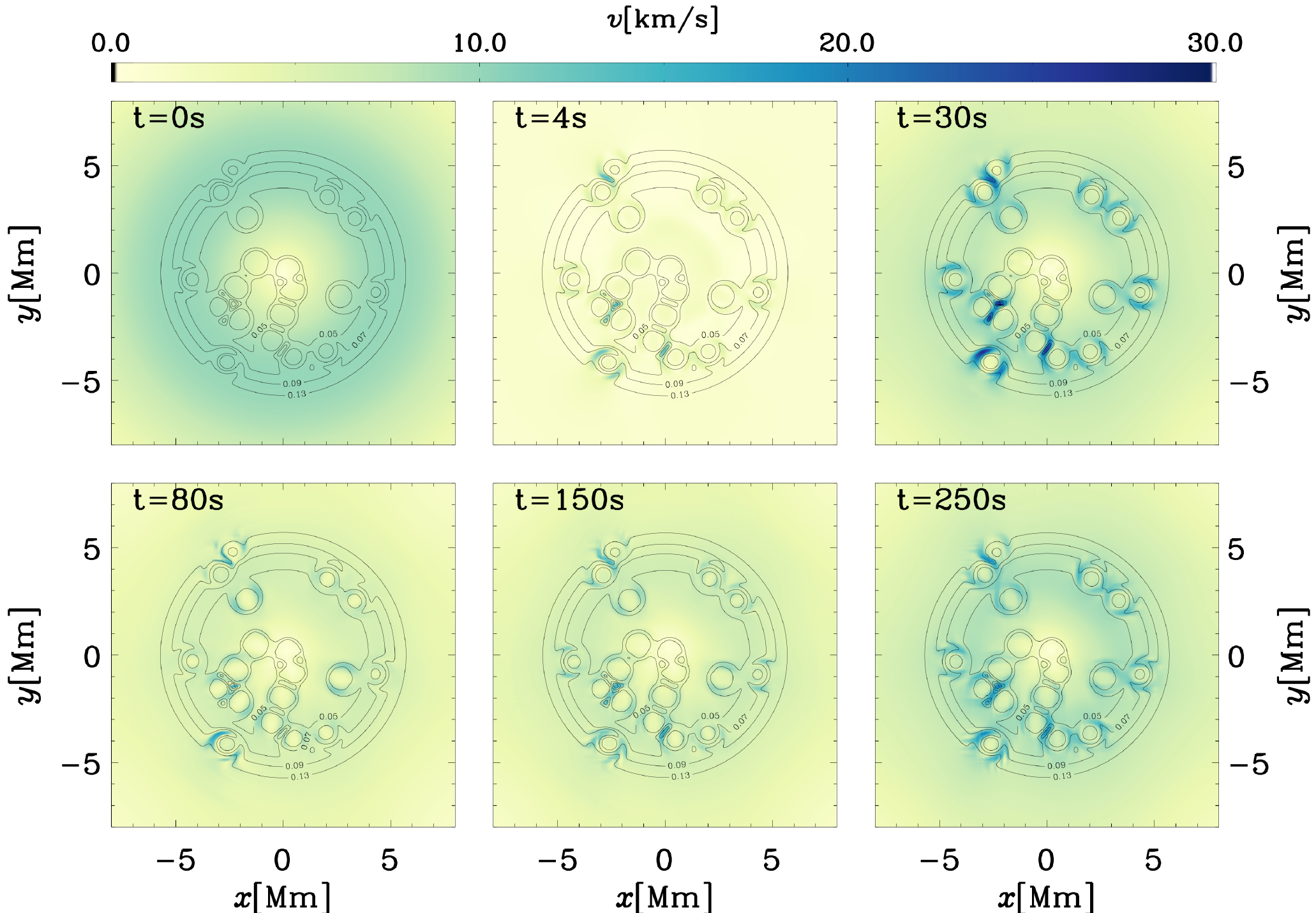}{0.9\textwidth}{}
	}
	\caption{
	   	Local velocity $v=\sqrt{v_x^2+v_y^2}$ at $z=L/2$ at different instants. Alfv\'en frequency $\omega_A$ is contoured at each instant. 
	}			
	\label{fig_res}
\end{figure}

\begin{figure}
	\centering
	\gridline{\fig{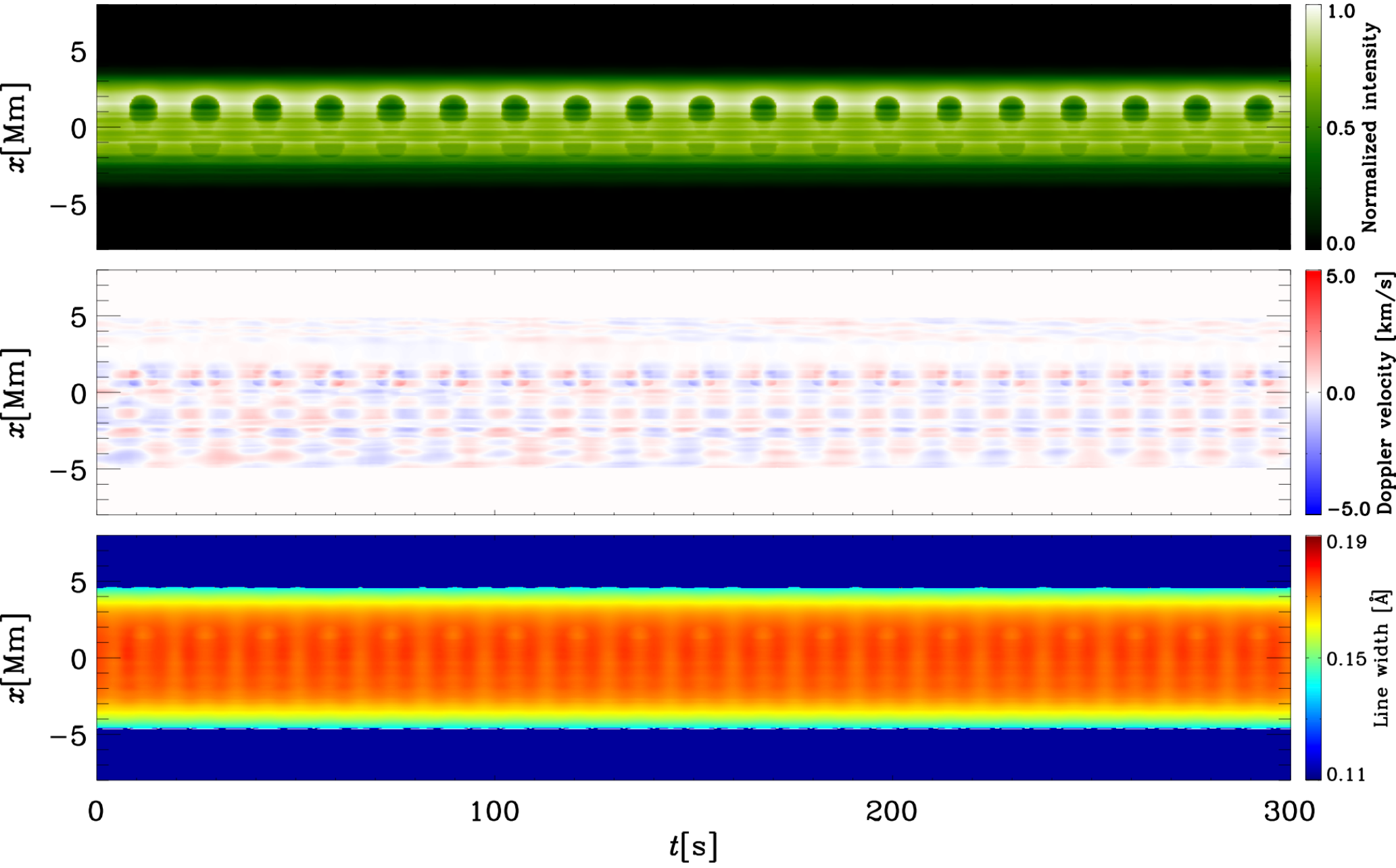}{0.8\textwidth}{}
	}
	\caption{
	   	Temporal evolution of synthetic intensity, Doppler velocity, and line width in the Fe  \uppercase\expandafter{\romannumeral 21} 1354 \AA~spectral line at $z=L/2$ with an LOS of $\pi/4$ with respect to the $x$-axis. 
	}			
	\label{fig_fomo_app}
\end{figure}

Another LOS angle of $\pi/4$ with respect to the $x$-axis is examined in Figure \ref{fig_fomo_app}.
The oscillation properties are similar to that in the main text with an LOS along the $y$-axis.
A similar periodicity as in Figure \ref{fig_fomo} can be seen here.
However,
the distribution of the intensity in the $x$-direction is different from the one in the main text.
We see the Doppler velocity has a slight drop in Figure \ref{fig_fomo_app}.
The line width maps are similar but slightly larger in Figure \ref{fig_fomo_app}.
All the differences presented here reveal the non-axisymmetry of our loop.

Note that the resolution in Figure \ref{fig_fomo} and Figure \ref{fig_fomo_app} remains the numerical one,
which is much higher than that of the real instruments.
So the fine structures in Figure \ref{fig_fomo} and Figure \ref{fig_fomo_app} can hardly be observed in reality.
Although the strands in our loop are not so thin and can be measured by many instruments that have a resolution of about 1 arcsec,
to recognize more structures due to the overlap of different strands, for instance,
may need a much higher resolution.
A similar discussion about the potential of future instruments can be found in \citet{2019ApJ...870...55G}.

\clearpage 

\clearpage 

\clearpage 

\clearpage 

\clearpage 

\clearpage 

\clearpage 

\clearpage 

\clearpage
\bibliographystyle{apj}
\bibliography{saus_FS.bbl}

\begin{figure*}

\end{figure*}

\end{document}